\documentclass[preprints,article,accept,moreauthors,pdftex]{Definitions/mdpi}
\firstpage{1} 
\pubvolume{1}
\issuenum{1}
\articlenumber{0}
\doinum{}
\pubyear{2021}
\copyrightyear{2020}
\datereceived{27 September 2021} 
\dateaccepted{19 October 2021 } 
\datepublished{} 
\hreflink{https://doi.org/} % If needed use \linebreak
\pdfoutput=1
\usepackage{aas_macros}

\Title{Multimodal analysis of Gravitational Wave signals and Gamma-Ray Bursts from binary neutron star mergers}
\TitleCitation{Title}

\Author{Elena Cuoco $^{1,2,3}$*\orcidA{}, Barbara Patricelli $^{1,3} \orcidB{}$, Alberto Iess \orcidC{} $^{4}$ and Filip Morawski $\orcidD{} ^{5}$}

\AuthorNames{Elena Cuoco, Barbara Patricelli, Alberto Iess and Filip Morawski}

\AuthorCitation{Cuoco, E.; Patricelli, B.; Iess, A.; Morawski, F.}
\address{%
$^{1}$ \quad European Gravitational Observatory (EGO); elena.cuoco@ego-gw.it (E.C.); barbara.patricelli@ego-gw.it (B.P.)\\
$^{2}$ \quad Scuola Normale Superiore (SNS); elena.cuoco@sns.it \\
$^{3}$ \quad INFN, Sezione di Pisa, I-56127 Pisa, Italy; barbara.patricelli@pi.infn.it \\
$^{4}$ \quad Universit\`{a} di Roma Tor Vergata, I-00133 Roma, Italy \\
$^{5}$ \quad Nicolaus Copernicus Astronomical Center, Polish Academy of Sciences, Bartycka 18, 00-716, Warsaw, Poland
}

\corres{Correspondence: elena.cuoco@ego-gw.it}

\abstract{A major boost in the understanding of the universe was given by the revelation of the first coalescence event of two neutron stars (GW170817)  and the observation of the same event across the entire electromagnetic spectrum. 
With 3rd Generation gravitational wave  detectors and the new astronomical facilities,  %(LSST, ELT, SKA, CTA, KM3NET), 
we expect many multi messenger events of the same type. We anticipate the need to analyse the data provided to us by such events, to fulfill the requirements of real-time analysis, but also in order to decipher the event in its entirety through the information emitted in the different messengers  using Machine Learning. 
We  propose a change  in the paradigm in the way we will do multi-messenger astronomy, using simultaneously the complete information generated by violent phenomena in the Universe.
What we propose is the application of a {\it multimodal  machine learning} approach to characterize these events.}

\keyword{gravitational waves; gamma-ray bursts; neutron stars; multi-messenger astronomy; multimodal analysis}

\begin{document}
%%%%%%%%%%%%%%%%%%%%%%%%%%%%%%%%%%%%%%%%%%
 
\section{Introduction}
 
The detection of gravitational waves (GWs) from the inspiral phase and coalescence of a pair of neutron stars (NS) on August 17th 2017 \cite{PhysRevLett.119.161101} and the following observations of the event in its electromagnetic (EM) counterparts  (see \cite{Abbott_2017_mm} and references therein) marked the beginning of multi-messenger astronomy  with GWs. \\ 
For the first time, we observed the coalescence of two NSs through GWs and EM radiation across the entire electromagnetic spectrum, thanks to the participation of more than 70 astronomical observatories to the EM follow-up campaign. 
Multi-messenger astronomy opens up new scenarios for the observation of the universe, new perspectives for the investigations of astronomical objects, but also new challenges for the way of extracting every information that these astrophysical events bring with them. The synergy between the information that only GWs can provide and the concomitant observations through other detectors of the EM  and neutrino counterparts can strongly accelerate our knowledge of the Universe. 
It is clear that multi-messenger astronomy discloses the need of new paradigms for data analysis and introduces new challenges for real-time analysis, and there are many efforts ongoing to face with them, which involve the use of machine learning techniques (see, e.g., \cite{2019A&A...631A.147N,2021arXiv210804166C,2020MNRAS.497.1320S,2021arXiv210612594D,2021PhLB..81536161K,2021MNRAS.508.1358S,2021PhRvD.104d2001B,2020PhLB..80335330K,2020ApJ...894L..25S,2020arXiv200803312G}).  Multimodal machine learning (MMML) analysis is efficiently applied in many fields of data analysis for the more inclusive interpretation of events where several modalities are concurrent, such as in a video with audio, or images with caption, or images, text and sound \cite{10.1109/TPAMI.2018.2798607}. To our knowledge, these techniques have never been applied in the interpretation of astrophysical data, where signals of different nature can be almost simultaneous. In this paper we introduce, for the first time, a multimodal machine learning analysis applied to astrophysical transient signals such as the case of GWs and Gamma-Ray Bursts (GRBs).

%\barbara{There is increasing interest also in multi-messenger real time analysis, and several tools have been developed for the identification of EM counterparts to GWs with ML techniques. One example is represented by El-CID \cite{2021arXiv210804166C}, that will allow the identification of kilonovae through the list of candidate counterparts to GWs in near real time after discovery.}

In Section \ref{sec:MMA} we will describe the importance of  multi-messenger astronomy; in section \ref{sec:MML} we will introduce the multimodal analysis and some example of application; in section \ref{sec:MMAMML} we will describe how we can implement the multimodal analysis for multi-messenger events and in \ref{sec:application} we will report a proof of concept for the the application of MMML to GW and GRB data. In the conclusion we will discuss about other possible applications. 

\section{Multi-messenger observations as a powerful tool to investigate the extreme universe}
\label{sec:MMA}
 The joint observation of GW170817 and its EM counterparts clearly demonstrated the enormous informative power of multi-messenger astronomy with GWs. For instance, the joint detection of GW170817 and GRB 170817A represents the first direct proof that NS-NS mergers are progenitors of short GRBs \cite{2017ApJ...848L..13A}. %In addition to this, the joint observation also allowed to infer some basic properties of short GRB jets (see, e.g., \cite{2019ApJ...880...55M}).
In addition to this, the observed time delay between the GW and the gamma-ray signal ($\sim$ 1.7 s) allowed us to put constraints on the difference between the speed of gravity and the speed of light, that  has been estimated to be between -3$\times 10^{-15}$ and $+7\times 10^{-16}$ times the speed of light \cite{2017ApJ...848L..13A}.
 After the joint detection of GW170817 and GRB 170817A, the release of a three-detector, well constrained GW skymap has been key for the identification of other EM counterparts \cite{2017ApJ...848L..12A}: this allowed us to get more insights into the physics of the source. For instance, the multi-wavelenght EM observations associated with GW170817 allowed us to infer some basic properties of short GRB jets. The temporal evolution of the X-ray \cite{2017Natur.551...71T} and radio \cite{2017Sci...358.1579H} light curves, together with the very low gamma-ray luminosity of GRB 170817A, suggested two possible scenarios: an off-axis GRB with a relativistic, structured jet or a ``cocoon'' emission from the relativistic jet shocking its surrounding non-relativistic material \cite{2017ApJ...848L..13A}. Subsequent very long baseline interferometry observations have been crucial to discriminate between these two scenarios: they allowed astrophysicists to put constraints on the size of the source and on its displacement, that were found to be consistent with a structured, relativistic jet \cite{2018Natur.561..355M,2019Sci...363..968G}.
 
 The detection of the optical/NIR counterpart to GW170817 (AT2017gf), first reported by \cite{2017GCN.21529....1C} and later by other teams (see \cite{Abbott_2017_mm} and references therein) allowed for the first time to identify the host galaxy of a GW event and led to the first spectroscopic identification of a kilonova \cite{2017Natur.551...67P,2017Natur.551...75S}, thus expanding our knowledge of heavy elements nucleosynthesis in the Universe. The joint GW and kilonova observation also allowed us to investigate in more detail the neutron star equation of state (EOS). For instance,  \cite{2018ApJ...852L..29R} found lower bound on the tidal deformability parameter through the interpretation of the UV/optical/IR counterpart of GW170817 with kilonova models, combined with new numerical relativity results; by combining this result with the constraints obtained with GW data alone, they shown that both extremely stiff and soft EOS are tentatively ruled out. \cite{2019MNRAS.489L..91C} presented a  Bayesian parameter estimation combining information from GW170817, GRB170817A and AT2017gfo, and with this analysis they were able to obtain multi-messenger constraints on the EOS and on the binary properties. More recently, \cite{2020Sci...370.1450D} performed a joint analysis of GW170817, GRB170817A and AT2017gfo, together with GW190425, and they combined these with previous measurements of pulsars using X-ray and radio observations, as well as and nuclear-theory computations, to put constraints on the EOS.
 A deeper knowledge of the EOS is fundamental also to understand which is the outcome of coalescing binary systems and therefore to constrain the short GRB central engine (see, e.g., \cite{2020MNRAS.499L..96P}).
 
 Finally, the estimate of the luminosity distance with GWs, together with the estimate  of the redshift obtained from the host galaxy, allowed us to estimate the Hubble constant with a totally new approach, independent from previous measurements \citep{2017Natur.551...85A}. Joint observations of GW and EM are thus a remarkable instrument for unveiling the physics of some of the most extreme phenomena in the Universe.

%%%%%%%%%%%%%%%%%%%%%%%%%%%%%%%%%%%%%%%%%%
\section{Artificial intelligence via multimodal inputs}
\label{sec:MML}
 Multimodal Machine Learning (MML) is a multidisciplinary research area that addresses some of the main objectives of Artificial Intelligence (AI) by incorporating and creating models that can process and link information from multiple modal inputs, differing in the representation (text, images etc.), dimensions (1-D, 2-D etc.) as well as input data sources.
 By considering data from multiple modalities, it is possible to take into account the complementary information among them, which in turn leads to more robust predictions that reflect patterns not available when working with individual modalities.

\begin{figure}[H]
\centering
\includegraphics[width=13cm]{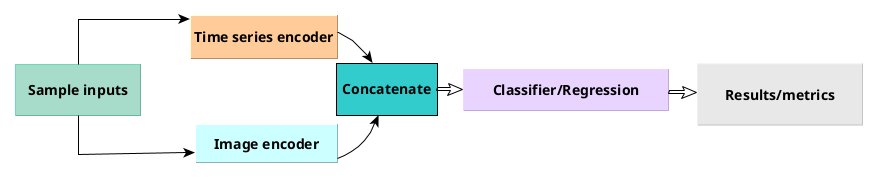}
\caption{Simple multimodal machine learning analysis workflow example.}
 
\label{fig:MMLworkflow}
\end{figure}

The multimodal approach is already used in a wide variety of artificial intelligence problems, such as use for virtual assistance, image captioning, question answering, etc.
One must consider the advantages in obtaining multiple input data characterizing the same event, as well as features extracted in different domain  decompositions. In this way we have the ability to capture otherwise unidentifiable details \cite{10.1109/TPAMI.2018.2798607}. 
In figure \ref{fig:MMLworkflow} we reported a schematic view of the general idea underlying the multimodal analysis.  The input samples contains different kind of signals  and representations which you can encode in a Deep Network analysis, concatenate and the later stage use for your classification/regression analysis.
%\begin{figure}[ht]

Machine learning (ML) and Deep Learning (DL) techniques, which have changed the way data is processed in recent years, have already been implemented into the Gravitational Wave Astronomy community \cite{MLreview}, thanks to the computational resources available to us in recent years and the implementation of algorithms that allow effective use of graphical processing units (GPUs). 
Recently, DL had been successfully applied even to multimodal machine learning problems, with the aim of learning useful joint representations in data fusion applications \cite{10.5555/3104482.3104569}. 
In \cite{Iess2020} we introduced a first approach to MML, based on the merging of deep learning pipeline outputs on 2 different kind of inputs: time series and images in the scheme which is called late fusion multimodal \cite{LateFusion}.

The challenge will be to apply these techniques also to data from different instruments with different outputs that characterize a multi-messenger event. We want to show how  MML could be used to process 1-D strain as well as 2-D spectrograms from GW detectors with sparse light-curves collected by astronomical telescopes in order to infer astrophysical information from the common sources. 
At the same time, even the ability to caption GW data with associated GRB event could help in fastly identify source parameters. 

%%%%%%%%%%%%%%%%%%%%%%%%%%%%%%%%%%%%%%%%%%
\section{From multi-messenger observations to multimodal analysis}
\label{sec:MMAMML}
In the next years, second generation GW interferometers (Advanced LIGO \cite{2015CQGra..32g4001L}, Advanced Virgo \cite{2015CQGra..32b4001A} and KAGRA \cite{2012CQGra..29l4007S,2013PhRvD..88d3007A}) 
will take data with increased sensitivity, and third generation GW detectors (such as the Einstein Telescope \cite{2017JApA...38...58A}) will become  operative; furthermore, many new telescopes will start taking data (e.g., CTA \cite{2013APh....43....3A}, LSST \cite{2019ApJ...873..111I}): we expect therefore an increase in the data rates and in the data complexity. In order to maximize the scientific return of future multi-messenger observations, there is the need to develop new approaches to analyse large streams of EM, GW and neutrino data taking into account the differences in instrument sensitivities, spatial and temporal coverage, data formats etc; furthermore, novel tools are needed to combine in an efficient way informations from multi-messenger observations, allowing us to infer the properties of the astrophysical sources and their environment. For instance, joint GW and EM observations can be used to put more stringent constraints on the tidal deformability parameter ($\Lambda$) and therefore on the EOS of NS (see, e.g., \cite{2018ApJ...852L..29R,2018MNRAS.480.3871C}; see also Section \ref{sec:MMA}). %, since both the phase of the GW signals and the kilonova light curve peak magnitude depend on $\Lambda$ (see, e.g., \cite{2018ApJ...852L..29R})\elena{Please, add references}. 
To do this, we need to perform a GW parameter estimation with an accurate family of template waveforms, as well as a detailed comparison between the EM observations and the existing kilonova theoretical models; informations obtained from the two messengers should then be combined in a consistent way to put constraints on the EOS. Such constraints, together with the estimates of the masses of the two NSs as obtained with GWs, could also help us to get insight into the outcome of the coalescence (that can be a NS or a BH) \cite{Dietrich1450}. In case of a coincident short GRB observation, the additional characterization of the X-ray afterglow emission will allow us to eventually probe the GRB-magnetar model, based on which  magnetars are the GRB central engine; such model has been revealed itself very successful in reproducing the observed properties for the sub-class of short GRBs showing an X-ray plateau \cite{2013MNRAS.430.1061R} and/or an extended emission \cite{2008MNRAS.385.1455M,2012MNRAS.419.1537B}, but a direct proof of this connection is still missing.% \elena{please, add reference}.

%In order to maximize the scientific return of future discoveries, there is the need to develop new approaches that combines EM and GW observations and allow us to infer the properties of the astrophysical sources and their environment (for instance, the NS tidal deformability, the GRB central engine...)
 In this work we propose a new paradigm for analyzing the multi-messenger data we will collect with next generation instruments.  

\begin{figure}[h!]
\centering
\includegraphics[width=13cm]{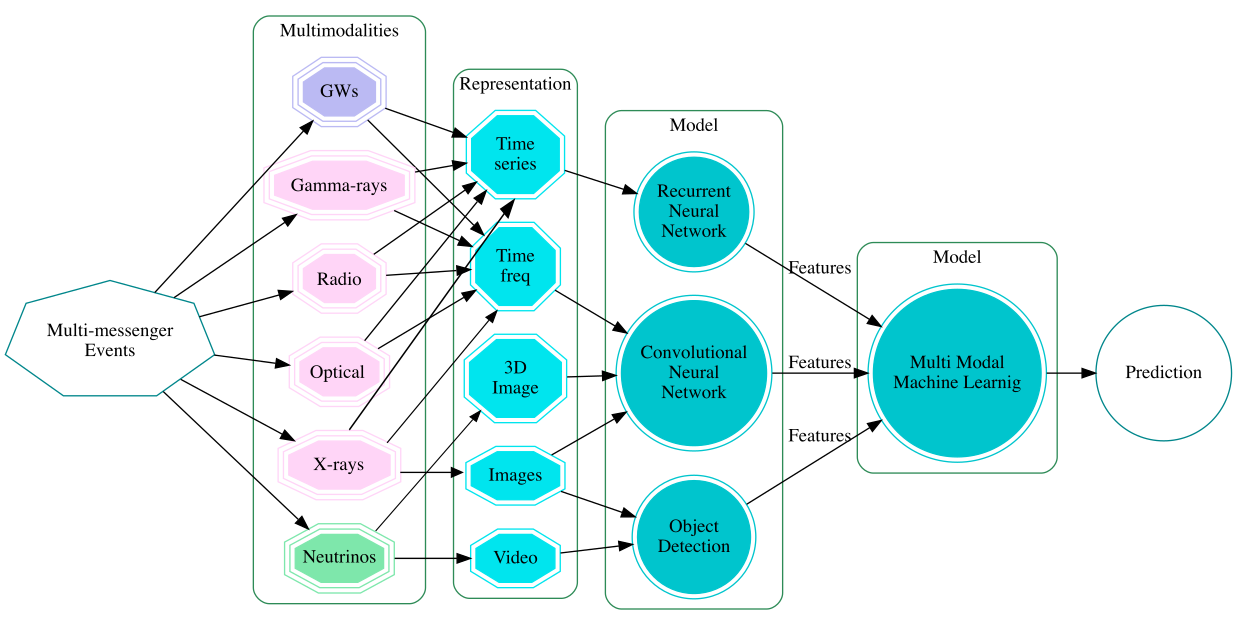}
\caption{An astrophysical phenomenon such as core-collapse supernova (CCSN), NS-NS or BH-NS coalescence (multi-messenger events) can manifest itself through different signal types, such as: gravitational waves, gamma-rays, X-rays, optical and radio emission, neutrinos. The different modalities have their own representations in different domains. By using DL and ML models, we can use the extracted features to do model prediction at a first stage.  At a later stage, we can use the predicted features by combining them in the global MML model.}
%An astrophysical phenomenon from CCSN, BNS or BHNS coalescence (multi-messenger events) can manifest itself through different signal types: gravitational wave, optical, radio, X-ray, GRB, neutrinos. The different modalities have its own representation in different domain. By using DL and ML models, we can use the extracted features to do model prediction at a first stage.  At a later stage, we can use the predicted features by combining them in the global MMML model.}
\label{fig:workflow}
\end{figure}

In figure \ref{fig:workflow} we sketched an example of multimodal analysis for multi-messenger events. We can approach the information of the diverse messengers through a dedicated pipeline, representing them in the best format for features extraction. We can analyze these representations using the best suited machine learning workflow to maximize the capability of prediction and at a final stage combine the output of extracted features in a multimodal model. 
We are depicting here a future vision, where in an open access environment we can analyze shared data using shared software on cloud systems \cite{allen2020escape}. 

%We show the application to the analysis of only data from gravitational wave experiments, where we simultaneously analysed the data in the temporal representation and the data in the image and wavelet decomposition.

\section{Application to astrophysical sources: the case of binaries of compact objects}

\label{sec:application}
 We decided to test our idea on a set of simulated short GRB light curves and associated GW events, with focus on NS-NS mergers. Specifically, we divided this task in three steps. 1) We simulated a sample of NS-NS merging systems populating the universe volume that can be explored with next generation GW interferometers. Specifically, we assigned to each NS-NS  system a luminosity distance that is randomly extracted from a uniform distribution in the range between 1 to 500 Mpc, to cover a realistic range of the matched filter Signal-to-Noise Ratio (SNR) - varying from 4 to 20. We then assumed that both components of the binary systems  have a mass distribution equal to a uniform distribution between 1 and 2.5 M$_\odot$; the two distributions were  assumed to be uncorrelated. The inclination angle ($\theta_i$) of the systems was chosen taking into account that GRB emission is collimated and these sources are typically detected when they are on-axis, i.e. with the jet pointing towards the observer. According to the few estimates currently available, the jet opening angle $\theta_j$ ranges from $3^\circ$ to $8^\circ$ (see, e.g., \cite{2014ApJ...780..118F} and references therein); we therefore restricted the range of possible values of $\theta_i$ to take this observation into account.
2) Following the approach presented in \cite{2016JCAP...11..056P}, we assumed that all the NS-NS mergers are associated with a short GRB and we simulated the associated high-energy afterglow light curves that could be observed by the LAT instrument onboard the {\sl Fermi} satellite \cite{2009ApJ...697.1071A}, using GRB 090510 \cite{2010ApJ...716.1178A} as a template. 
3) We simulated the GW signal associated with the NS-NS mergers using the TaylorF2 waveform model \cite{2009PhRvD..80h4043B}. We simulated the noise data for a GW detector such as the Einstein Telescope \cite{2011CQGra..28i4013H}, where we injected the NS-NS merger GW signals. The simulations were performed using the  pyCBC library \cite{pycbc}. 
\begin{figure}[h!]
\begin{minipage}{.6\textwidth}
\centering
    \hspace*{-5cm}
    \includegraphics[width=0.6 \textwidth]{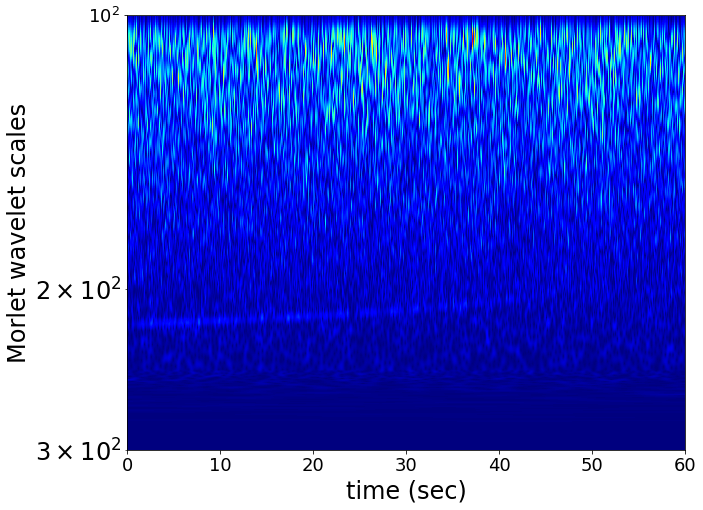}
\end{minipage}%%
\begin{minipage}{.6\textwidth}
\centering
    \hspace*{-13.0cm}
    \includegraphics[width=0.6 \textwidth]{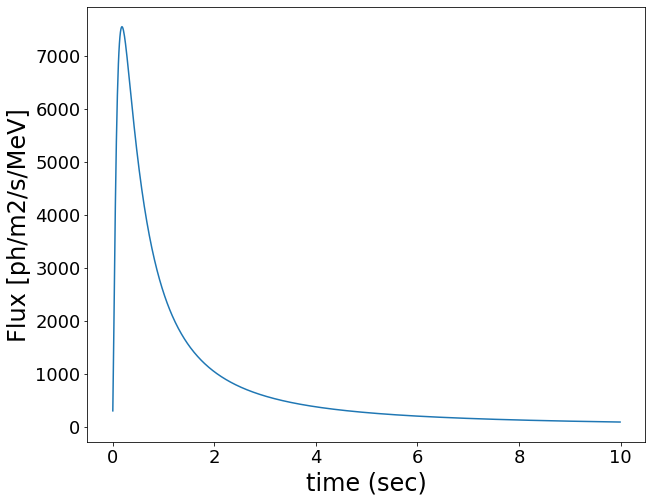}
\end{minipage}
  \caption{Sample binary neutron star with component masses 2.4$M_\odot$  and 1.8M$_\odot$ used as input to the ML model, at a distance of 283\,Mpc. The gravitational wave signal ({\textit left}) and the GRB light curve ({\textit right}) both contribute to the prediction. We turned off the x/y label (time-frequency information) to feed the CCN-2D network with images. }
  \label{fig:samples}
\end{figure}

We propose a MML pipeline consisting of two Convolutional Neural Networks (CNN) concatenated at the ouput in order to estimate the redshift of the GRB and GW sources. To convert luminosity distances in redshifts, we use the cosmological parameters in \cite{2016A&A...594A..13P}. A 2-D CNN takes the time-frequency image representation of the GW signal as input. The images are built from the detector strain time series containing the injected NS-NS GW signals. As a first step, the strain is whitened in time domain by means of an Auto-Regressive (AR) model \cite{cuoco-whitening} to remove the stationary noise component. The 60 seconds long segments containing the chirp signals are then converted into a time-frequency representation based on the continuous wavelet transform, using Morlet wavelets 
(using ssqueezepy library \footnote{https://github.com/OverLordGoldDragon/ssqueezepy}), see \cite{Thakur2011,Thakur2013,Daubechies2011}. %\cite{ssqueezepy, Thakur2011,Thakur2013,Daubechies2011}.
Finally the images are compressed to dimensions in pixel of $128\times256$. For GRB lightcurves, we used the data in time domain. The length of GRB simulated data was kept up to 1000 points, where most of the information was encoded\footnote{We used a time resolution of 0.01 s, so 1000 points correspond to a time interval of 10 s}. Examples of the chosen representations are shown in Fig. \ref{fig:samples}: on the left side the image for GW signal, on the right side the time domain light curve data.

\begin{figure}[h!]
\centering
\includegraphics[width=10.5cm]{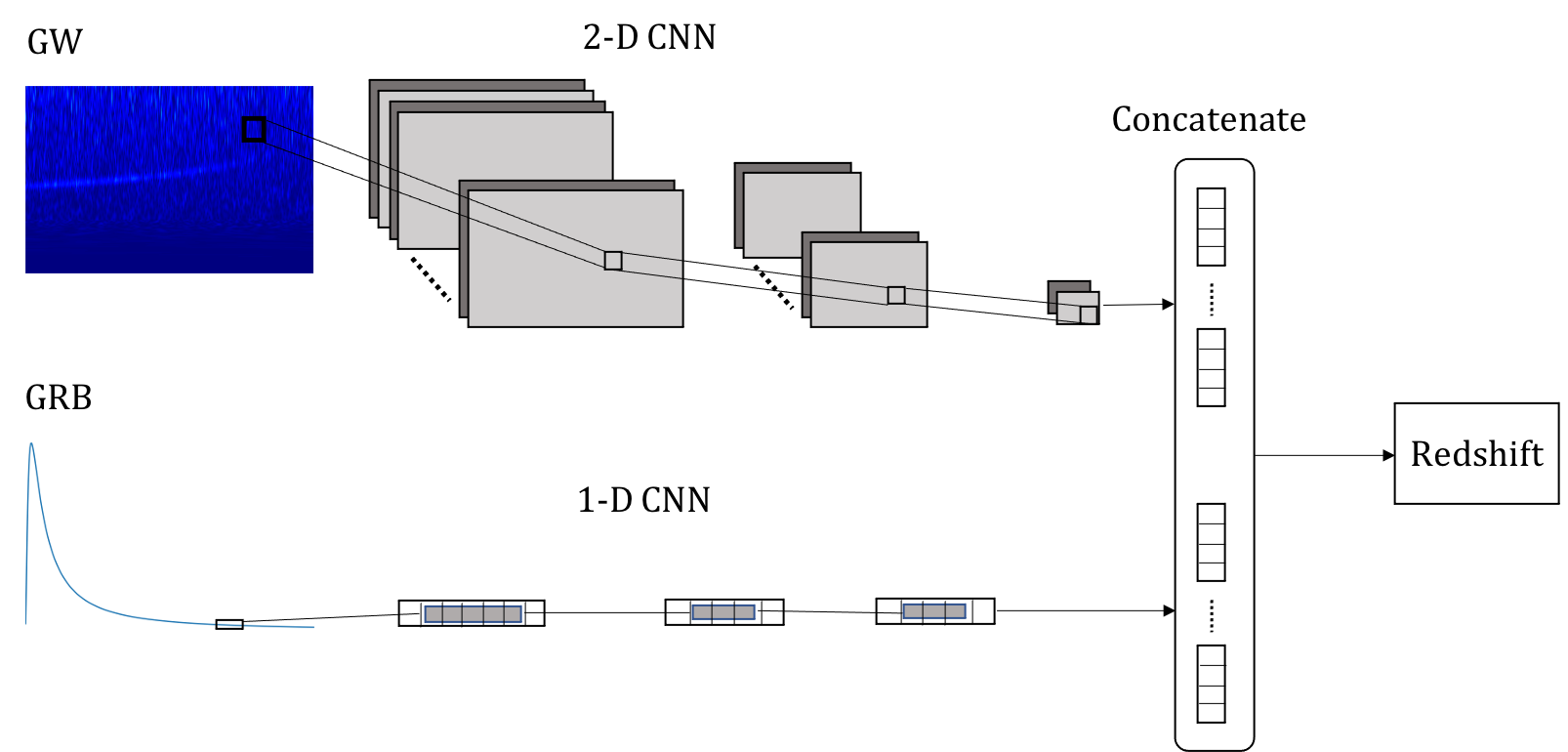}
\caption{Illustration of the multimodal machine learning model adopted in the analysis to compute redshifts of joint GW and GRB sources. The network reads two different types of data, images and time series, to tackle the regression problem.}

\label{fig:MMmodel}
\end{figure}
\begin{figure}[h!]
    \centering
    \includegraphics[width=12cm]{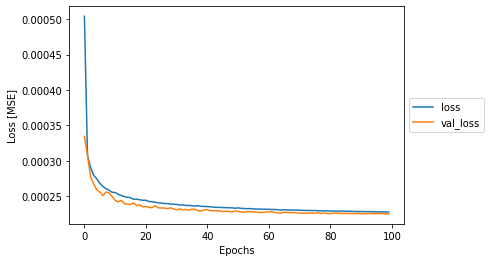}
    \caption{Training (in blue) and validation (in orange) MSE loss over epochs.}
    \label{fig:loss}
\end{figure} 
 The CNN processing GW data consists of $5$ layers with the following number of filters: $64, 32, 16, 16, 32$ and kernels of dimensions ($3,3$). After every convolutional layer, we applied maxpooling with kernels ($2,2$). The 1-D CNN processing light-curves consists of $3$ layers with the following number of filters: $80, 40, 40$ and kernels: $5, 3, 3$. Also in this case after every convolutional layer we applied maxpooling with kernel $2$. After the last layers, both CNN were flattened and concatenated as one. The concatenated output is fed to a fully connected layer which outputs the prediction for the source redshift. All layers have a ReLU activation function, with the exception of the final layer with a linear activation. The MML model was trained on the train data set using the Mean Squared Error (MSE)  loss function and Adam optimizer with learning rate of $\alpha=0.001$ and a learning rate decay of 0.066667. The model is summarised in Fig. \ref{fig:MMmodel}. For training and evaluation, the dataset was divided with the following scheme: $70\%$ training set, $20\%$ test set, $10\%$ validation set. Training was carried out with a batch size of 16 samples for 100 epochs. The total number of trainable parameters is 61881. As a pre-processing step, we applied minmax scaling to the inputs. Shuffling is also applied.
 
In Fig. \ref{fig:loss} we report the mean squared error loss evolution over the training epochs for both the train and validation sets. It can be observed that the MSE converged well over the 100 epochs algorithm, and that there is no overfitting.
\begin{figure}[h!]
    \centering
    \includegraphics[width=10cm]{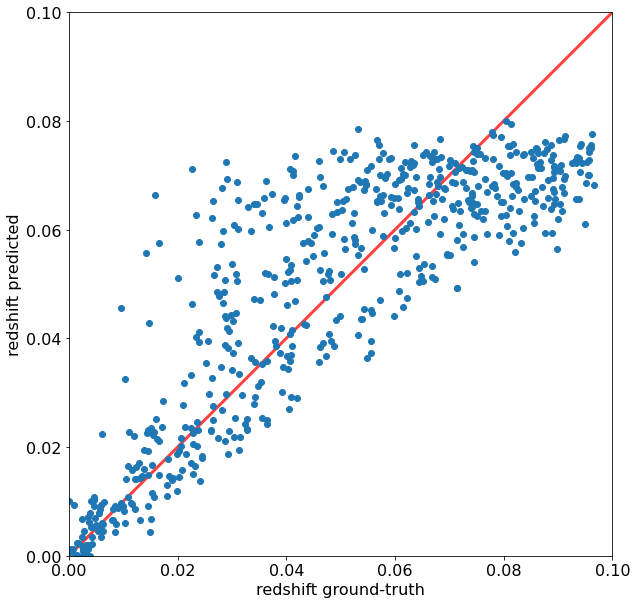}
    \caption{Predicted redshifts versus the true redshifts of the simulated sources (blue points); the red solid line represents the expectations in case that the predicted redshifts exactly match the true redshifts.}
    \label{fig:pred-truth}
\end{figure}
In Fig. \ref{fig:pred-truth} we show the comparison between the predicted redshifts and the true redshifts of the simulated sources on the test data set. It can be seen that, for the lowest values of redshifts, the predictions are in good agreement with the real values, while for the highest redshifts the scatter around the line $y=x$  increases and the predicted values are  typically lower than the true ones. This is partially due to the fact that, for the most distant sources, the GW signal has low SNR value. Also, we used a limited dataset, and this could have affected our results. However, with this work we only wanted to investigate the feasibility of multi-modal ML analysis; a more detailed study, including a larger simulated dataset and possibly real EM and GW data, will be presented elsewhere.

\begin{figure}[h!]
 
    \includegraphics[width=0.6 \textwidth]{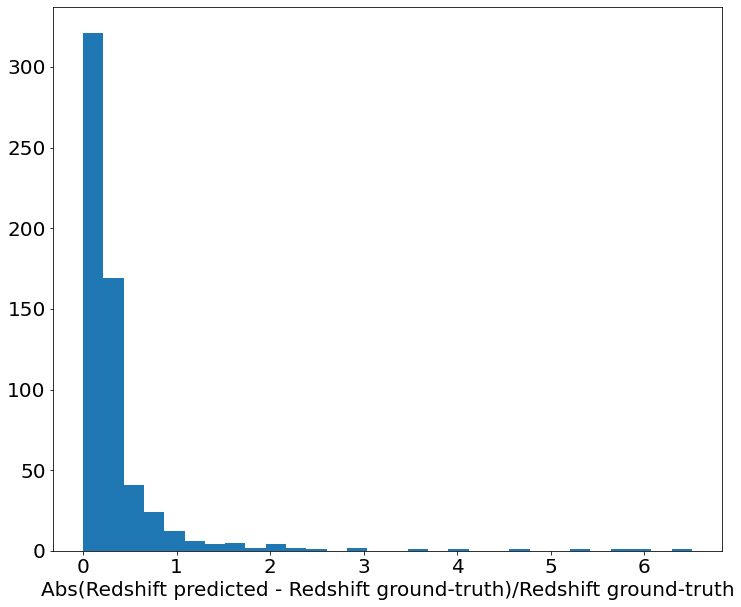}
     
    \caption{Histogram of the relative error in the estimated distance with respect true values.} \label{fig:histo}
\end{figure}

In Fig. \ref{fig:histo} we show the histogram of the relative difference between the predicted redshifts and the true redshifts . It can be seen that the histogram has a peak on zero, meaning that for the majority of the simulated sources the algorithm allowed to correctly estimate the redshift.

%%%%%%%%%%%%%%%%%%%%%%%%%%%%%%%%%%%%%%%%%% 
\section{Outlook and Perspective}
\label{sec:conclusion}
Our team's innovative Multimodal Machine Learning based   approach consists in analyzing each multi-messenger event simultaneously in its entirety, through the information we can gather from GWs or electromagnetic or neutrino emissions, which will be the inputs for a single MMML pipeline.
This allows us to increase the information we can extract, thanks to the concomitant analysis of every piece of information.
Just as it is easier for us to understand language when we associate sound to video, or to understand people's mood by analysing text and images, we will apply multimodal analysis techniques to astrophysical transient events. 

This is a new paradigm for analysing the data we will collect in future astro and particle experiments. We tested the idea on a data set we created simulating both GW and GRB emission, just to setup a proof of concept. We built a MML pipeline taking as input the images obtained through a time frequency representation of whitened GW data for Einstein Telescope detector, and the simulated GRB light curves for a Fermi-like detector.  We want to underline that the work here presented is just an example of a basic application of multimodal analysis for multi-messenger events, based only on GW strain data and gamma-ray GRB light curves, with a simple neural network architecture.
The results even if preliminary are very encouraging and we plan to continue with our approach on real data or including other input messengers. We will implement a more sophisticated and comprehensive analysis, by adding data in different formats, at different levels (from the raw data to the high-level data) and related also to other messengers (e.g., neutrinos and photons at other wavelengths). For instance, we want to include into our framework raw data from Imaging Atmospheric Cherenkov Telescopes (IACTs) such as CTA: 2-D images that represent the tracks  left in the telescope's camera by showers of particles that can be induced, for example, by very-high-energy (VHE, E $>$ 100 GeV) gamma rays emitted by GRBs. %and/or charged cosmic rays}
At the same time we plan to optimize the analysis workflows using feature extraction and engineering, data balance and augmentation for less represented class of events and working with deeper networks. 
It is worth to emphasize  that this approach we are proposing represents a new challenge for data scientists and astrophysicists, even in anticipation of the larger number of events we will be able to detect with third-generation GW detectors.

%%%%%%%%%%%%%%%%%%%%%%%%%%%%%%%%%%%%%%%%%%
%\section{Patents}

%This section is not mandatory, but may be added if there are patents resulting from the work reported in this manuscript.

%%%%%%%%%%%%%%%%%%%%%%%%%%%%%%%%%%%%%%%%%%
\vspace{6pt}

\dataavailability{The data presented in this study are available on request from the corresponding author.}
%%%In this section, please provide details regarding where data supporting reported results can be found, including links to publicly archived datasets analyzed or generated during the study. Please refer to suggested Data Availability Statements in section ``MDPI Research Data Policies'' at \url{https://www.mdpi.com/ethics}. You might choose to exclude this statement if the study did not report any data.} 

\acknowledgments{This article/publication is based upon work from COST Action CA17137, supported by COST (European Cooperation in Science and Technology). }

\conflictsofinterest{The authors declare no conflict of interest.}
%%%Declare conflicts of interest or state ``The authors declare no conflict of interest.'' Authors must identify and declare any personal circumstances or interest that may be perceived as inappropriately influencing the representation or interpretation of reported research results. Any role of the funders in the design of the study; in the collection, analyses or interpretation of data; in the writing of the manuscript, or in the decision to publish the results must be declared in this section. If there is no role, please state ``The funders had no role in the design of the study; in the collection, analyses, or interpretation of data; in the writing of the manuscript, or in the decision to publish the~results''.} 

%% Optional
%%%\sampleavailability{Samples of the compounds ... are available from the authors.}

%%%%%%%%%%%%%%%%%%%%%%%%%%%%%%%%%%%%%%%%%%
%% Only for journal Encyclopedia
%\entrylink{The Link to this entry published on the encyclopedia platform.}

%%%%%%%%%%%%%%%%%%%%%%%%%%%%%%%%%%%%%%%%%%
%% Optional
%%%\abbreviations{Abbreviations}{
%%%The following abbreviations are used in this manuscript:\\

%%%\noindent 
%%%\begin{tabular}{@{}ll}
%%%MDPI & Multidisciplinary Digital Publishing Institute\\
%%%DOAJ & Directory of open access journals\\
%%%TLA & Three letter acronym\\
%%%LD & Linear dichroism
%%%\end{tabular}}

%%%%%%%%%%%%%%%%%%%%%%%%%%%%%%%%%%%%%%%%%%
%% Optional
\appendixtitles{no} % Leave argument "no" if all appendix headings stay EMPTY (then no dot is printed after "Appendix A"). If the appendix sections contain a heading then change the argument to "yes".
%%%\appendixstart
%%%\appendix
%%%\section{}
%%%\subsection{}
%%%The appendix is an optional section that can contain details and data supplemental to the main text---for example, explanations of experimental details that would disrupt the flow of the main text but nonetheless remain crucial to understanding and reproducing the research shown; figures of replicates for experiments of which representative data are shown in the main text can be added here if brief, or as Supplementary Data. Mathematical proofs of results not central to the paper can be added as an appendix.

%%%\begin{specialtable}[H] 
%%%\small
%%%\caption{This is a table caption. Tables should be placed in the main text near to the first time they are~cited.\label{tab2}}
%%%\begin{tabular}{ccc}
%%%\toprule
%%%\textbf{Title 1}	& \textbf{Title 2}	& \textbf{Title 3}\\
%%%\midrule
%%%Entry 1		& Data			& Data\\
%%%Entry 2		& Data			& Data\\
%%%\bottomrule
%%%\end{tabular}
%%%\end{specialtable}

%%%\section{}
%%%All appendix sections must be cited in the main text. In the appendices, Figures, Tables, etc. should be labeled, starting with ``A''---e.g., Figure A1, Figure A2, etc. 

%%%%%%%%%%%%%%%%%%%%%%%%%%%%%%%%%%%%%%%%%%
\end{paracol}

%%%%%%%%%%%%%%%%%%%%%%%%%%%%%%%%%%%%%%%%%%
% To add notes in main text, please use \endnote{} and un-comment the codes below.
%\begin{adjustwidth}{-5.0cm}{0cm}
%\printendnotes[custom]
%\end{adjustwidth}
%%%%%%%%%%%%%%%%%%%%%%%%%%%%%%%%%%%%%%%%%%
\reftitle{References}


%\externalbibliography{yes}
%\bibliography{your_external_BibTeX_file}

%=====================================
% References, variant B: internal bibliography
%=====================================
\bibliography{references}

\end{document}